\documentclass[doublecol]{epl2} 

\usepackage{graphicx}
\usepackage{bm}
\usepackage[english]{babel}
\usepackage[latin1]{inputenc} 
\usepackage{graphicx}
\usepackage{epstopdf}
\usepackage{amsfonts}
\usepackage{color}
\usepackage{epsfig}
\usepackage{mathrsfs}
\usepackage{amsmath,amssymb}

\renewcommand{\(}{\left(}
\renewcommand{\)}{\right)}

\title{Scaling and Universality in River Flow Dynamics}
\shorttitle{Scaling and Universality in River Flow Dynamics}

\author{M. De Domenico \and V. Latora}

\institute{                    
   Laboratorio sui Sistemi Complessi, Scuola Superiore di Catania - Via Valdisavoia 9, 95123 Catania, Italy, EU and\\
    Dipartimento di Fisica e
    Astronomia, Universit\`a di Catania, and INFN - Via S. Sofia 64, 95123
    Catania, Italy, EU
}

\abstract{ We investigate flow dynamics in rivers characterized by basin areas and daily mean discharge spanning different orders of magnitude. We show that the delayed increments evaluated at time scales ranging from days to months can be opportunely rescaled to the same non-Gaussian probability density function. Such a scaling breaks up above a certain critical horizon, where a behavior typical of thermodynamic systems at the critical point emerges. We finally show that both the scaling behavior and the break up of the scaling are universal features of river flow dynamics.}

\begin{document}

\selectlanguage{english}

\maketitle


One of the grand challenges in hydrology is understanding the real nature of river flow dynamics. In fact, fluctuations in river flow are the result of complicated interactions among several external factors, such as climatic changes and landscape characteristics. Several attempts have been made to capture either stochastic or deterministic signatures in the dynamics of fluctuations at different time and space scales. However, a definitive comprehension of the phenomenon is still missing.

River flow clearly shows long-range correlations and anomalous scaling typical of multifractal processes \cite{tessier1996multifractal, pandey1998multifractal}. The Bramwell-Holdsworth-Pinton (BHP) distribution, describing fluctuations of global quantities in correlated turbulent and critical phenomena \cite{bramwell1998universality,bramwell2000universal}, has been recently proposed to explain universal scaling signatures for fluctuations in river level \cite{bramwell2002universal} and river flow \cite{dahlstedt2005fluctuation}, after rescaling time series with respect to the seasonal mean and variance. However, the nature of the fluctuations and multifractal behavior can not be uniquely related to a stochastic or a chaotic deterministic dynamics, because the observed dynamics changes from a less complex (deterministic) to a more complex (stochastic) one by varying the scale of aggregation \cite{porporato1997nonlinear, pasternack1999does, lisi2001chaotic, islam2002characterization, montanari2004stochastic}. Thus, apparently controversial results emerge from the analyses of river flow databases, because of the observed transitions from determinism to stochasticity with increasing time scale (see \cite{Sivakumar09} and Ref.\,therein). 

In this Letter we present an analysis of river flow fluctuations based on a method originally introduced for the investigation of turbulent flows and complex hierarchical systems. The key point of the analysis is to consider the time series of increments, a set of $N$ new time series, each constructed as the difference between the original time series, locally detrended, and a copy of it delayed by $n$ ($n=1,2,...,N$) time units. By investigating the probability density functions (PDFs) of the increments, we find a scaling of the non-Gaussian curves obtained at different time scales, that is typical of processes with anomalous diffusion. Moreover, we reveal the universality of such a scaling by comparing several different rivers. Our results suggest the presence of turbulence-like behavior in the fluctuations of river flows at any time scale, from daily to monthly, up to a certain time horizon, above which a behavior typical of thermodynamic systems at the critical point emerges.

\begin{figure}[!h]
  \begin{center}

	\includegraphics[angle=0, scale=0.3]{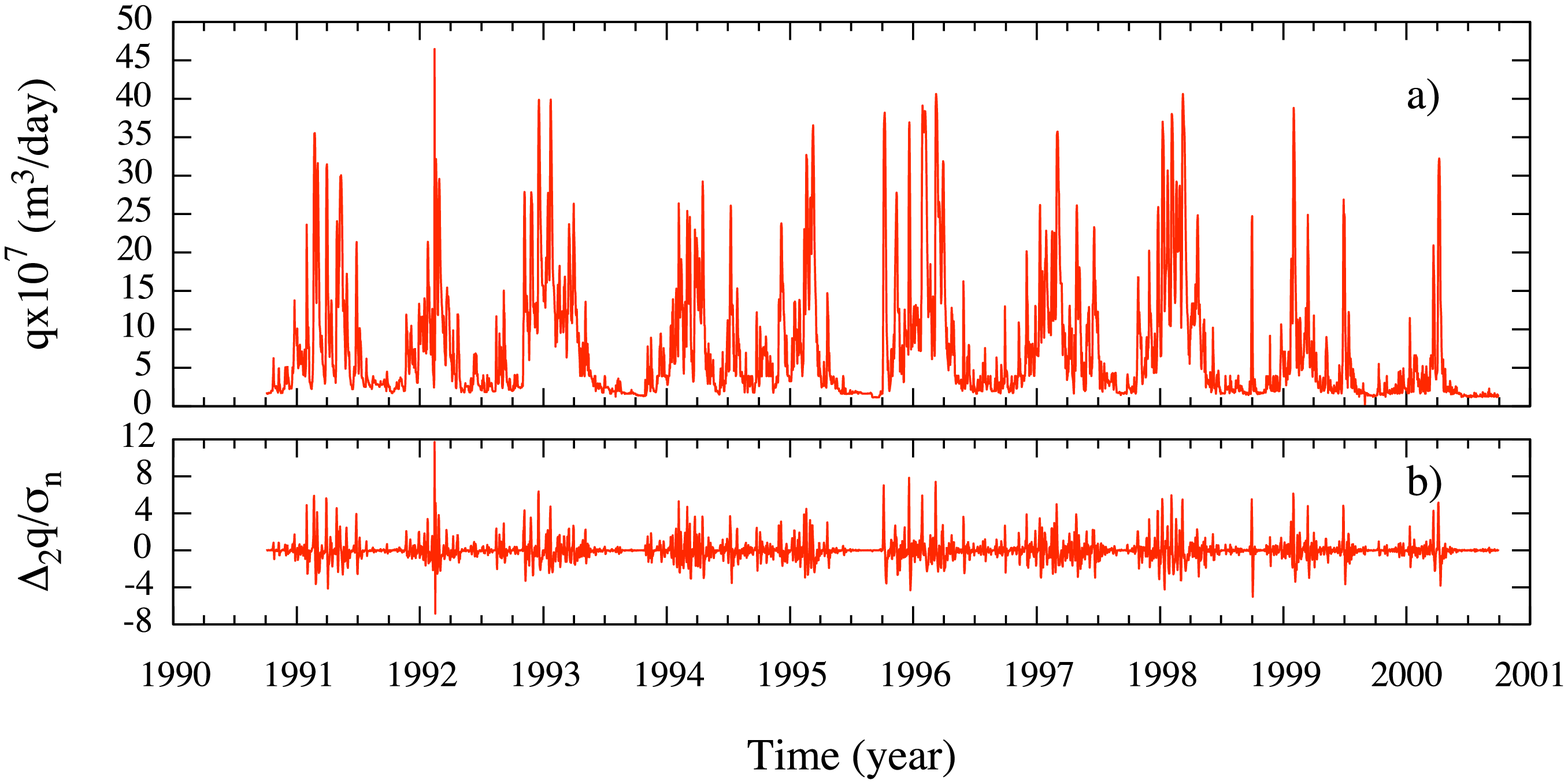}
      \caption{{(\emph{a}) Historical time series corresponding to ten years of daily discharge for the Alabama river as measured at the Millers Ferry station, AL. (\emph{b}) Delayed increments of the detrended time series for a time delay $n=2$ days.}}
	\label{fig-river}
  \end{center}
\end{figure}

We analyze historical time series of the average daily discharge as mesaured at several sites in the northeastern gulf of Mexico (NEGOM) region (USA), that covers a large area corresponding to a longitude ranging from $-91^{\circ}$ to $-82^{\circ}$ and a latitude ranging from $28^{\circ}$ to $33^{\circ}$. Moreover, we consider several other rivers around the world, from Asia, Europe and Australia. The list of all sites used for this study is shown in Tab.\,\ref{tab-rivers}: time series of different size, measured in different epochs, and corresponding to stations with different physical features, as territory and drainage area, are present. A representative example of the apparently irregular behavior of river flow,  is given in Fig.\,\ref{fig-river}(a), where ten years of measurements of the Alabama river discharge, at the Millers Ferry station (AL), are shown. In Fig.\,\ref{fig-river}(b) we report the series of delayed and detrended increments for the scale $n=2$ days, obtained as successively explained in the text.

\begin{table*}[!htb]
\centering\small
\begin{tabular}{llrl}
\hline
\hline
$\kappa$ & \textbf{Station name} & \textbf{Mean discharge} & \textbf{Historical data}\\
 &  & \textbf{($m^{3}/\text{day}$)} & \textbf{(from - to)}\\
\hline
\hline
1  &  Alabama (Millers Ferry, AL)  & 80300300  & 10/1969 - 09/2000  \\
2  &  Apalachicola (Sumatra, FL)  & 63265600  &  10/1977 - 04/2001  \\
3  &  Tombigbee (Demopolis, AL)     & 57410700  &      10/1927 - 04/2001  \\
4  &  Pascagoula (Graham Ferry, MS) &  25518700  & 10/1993 - 04/2001  \\
5  &  Aucilla (Scanlon, FL)        &   17600000  &   08/1976 - 09/1998  \\
6  &  Choctawhatchee (Bruce, FL)  & 17101400 & 10/1930 - 04/2001  \\ 
7  &  Suwannee (Branford, FL)     &  16804200  &  10/1930 - 04/2001  \\
8  &  Pearl (Monticello, MS)       &  16548000   &  10/1938 - 04/2001   \\
9  &  Escambia (Century, FL)    & 15348000  &  10/1934 - 04/2001  \\
10  &  Escambia (Molino, FL)       & 14333400  &    10/1982 - 03/2000  \\
11  &  Perdido (Barrineau Park, FL)  & 1903000 &  10/1940 - 09/2000  \\
12  &  Ochlockonee (Bloxham, FL)   & 4148310 &  10/1925 - 04/2001  \\
13  &  Yellow (Milligan, FL)       &   2742960   &   10/1937 - 04/2001  \\
14  &  Blackwater (Baker, FL)      &   832902 &    10/1949 - 04/2001  \\
15  &  Steinhatchee (Cross City, FL) & 763213 & 10/1949 - 01/2000  \\
16  &  Econfina (Perry, FL)       &  342289   &  10/1949 - 09/2000  \\
17  &  Fenholloway (Foley, FL)    & 124025  & 10/1955 - 09/1999  \\
18  &  Fish Creek (Ketchikan, AK) & 1037462 & 12/1938 - 12/2009\\
19  &  Inabòn (Real Abajo, Puerto Rico) & 45771 & 07/1964 - 12/2009 \\
20  &  Thames (Kingston, UK) & 5634413 & 01/1883 - 12/2009 \\
21  &  Severn (Bewdley, UK) & 5252122 & 04/1921 - 12/2009\\
22  &  Tana (Polmak, Norway) & 14535168 & 02/1946 - 12/1999 \\
23  &  Namsen (Bjornstad, Norway) & 3266237 & 10/1934 - 12/1997 \\
24  &  Verdalsvassdraget (Grunnfoss, Norway) & 3297774 & 11/1951 - 12/2000\\
25  &  Chiang Mai (Ping River, Thailand) & 5016590 & 01/1921 - 12/1999\\
26  &  Nakhon Sawan (Chao Phraya, Thailand) & 60307717 & 01/1956 - 12/1999 \\
27  &  Shuya (Shuyeretskoye, Russia) & 739335 & 01/1948 - 12/1988 \\
28  &  Tan-Shui (Fu-Shan, Taiwan) & 154554910 & 01/1953 - 12/1993 \\
29  &  Pei-Nan Chi (Yen-Ping, Taiwan) & 308124037 & 01/1956 - 12/1993 \\
30  &  Nan-Ao (Shan-Chiao, Taiwan) & 51564903 & 01/1954 - 12/1993\\
31  &  Cho-Shui (Tung-Tou, Taiwan) & 165580533 & 01/1956 - 12/1993\\
32  &  Leigh (Mount Mercer, Australia) & 134463 & 10/1956 - 12/2009 \\
33  &  Broken (Moorngag, Australia) & 188197 & 12/1973 - 12/2009 \\
34  &  Barwon (Inverleigh, Australia) & 290761 & 04/1966 - 12/2009\\
\hline
\hline
\end{tabular}
\caption{{List of rivers (at the specified station) in the northeastern Gulf of Mexico (NEGOM) region ($\kappa=1-17$) and around the globe ($\kappa=18-34$), whose historical average daily discharge data is used for the analyses in this work. Data is provided from U.S. Geological Survey ($\kappa=1-19$), National River Flow Archive ($\kappa=20-21$), Norwegian Water Resources and Energy Directorate ($\kappa=22-24$), Royal Irrigation Department-GAME-T2 Data Center ($\kappa=25-26$), RosHydromet ($\kappa=27$), Pacific Rim Streamflow Data Set ($\kappa=28-31$), Department of Natural Resources and Environment-Victorian Water Resources ($\kappa=32-34$).}}
\label{tab-rivers}
\end{table*}

\begin{figure}[!h]
  \begin{center}

     	\includegraphics[angle=0, scale=0.3]{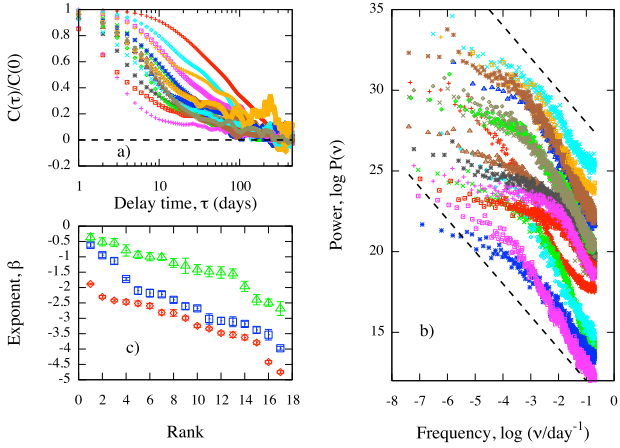}
      \caption{{Daily discharge historical data of the $\kappa-$th station ($\kappa=1,2,...,17$) in Tab.\,\ref{tab-rivers}. (\emph{a}) Autocorrelation function vs. delay time of deseasonalized time series; (\emph{b}) Power spectrum vs. frequency: dashed lines correspond to straight lines with slope -2; (\emph{c}) Rank-ordered scaling exponents ($P_{\kappa}(\nu)\sim \nu^{\beta_{\kappa}}$) corresponding to the slopes of the linear regions in Fig.\,\ref{fig-power}(b), for $\log(\nu/\text{day}^{-1})\in[-2,-1]$ (diamonds), $[-3,-2]$ (squares) and $[-4,-3]$ (triangles).}}    
      \label{fig-power}

  \end{center}
\end{figure}

We firstly investigate the average daily flow $q_{\kappa}$, $\kappa=1,2,...,34$, of all stations in Tab.\,\ref{tab-rivers}. By labelling with $q_{\kappa}(t)$ the measurement at time $t$, $t=1,2,..., N_{\kappa}$ days, at the $\kappa-$th station, we inspect both the autocorrelation function and the power spectrum of each data set. The former is defined from the autocovariance
\begin{eqnarray}
C_{\kappa}(\tau) = \frac{1}{N_{\kappa}-\tau}\sum_{t=1}^{N_{\kappa}-\tau}[\(q_{\kappa}(t)-\langle q_{\kappa}\rangle\)\(q_{\kappa}(t+\tau)-\langle q_{\kappa}\rangle\)]\nonumber
\end{eqnarray}
as $C_{\kappa}(\tau)/C_{\kappa}(0)$, and quantifies the amount of correlation among measurements delayed by an interval of amplitude $\tau$ (the variance of the series corresponds to $C_{\kappa}(0)$). The delay $\tau^{\star}$ such that the autocorrelation reaches zero for the first time, is known as decorrelation time, and estimates the maximum range of dependence among samples. The power spectrum can be defined through the Wiener-Khinchin theorem as 
$P_{\kappa}(\nu)=\mathcal{F}[C_{\kappa}(\tau)/C_{\kappa}(0)]$, where $\mathcal{F}$ is the Fourier transform, which quantifies the amount of correlation among samples in the frequency domain. In order to reduce fluctuations at higher frequencies, the power spectrum of a single data set is obtained by dividing $q_{\kappa}$ into 16 non-overlapping subsamples, by calculating the periodogram for each subsample, and by estimating the ensemble average. As a representative example, in Fig.\,\ref{fig-power}(a) is shown the autocorrelation function for all rivers corresponding to $\kappa=1,2,...,17$. Time series have been previously deseasonalized  by rescaling with respect to the seasonal mean and variance, as in Refs. \cite{dahlstedt2005fluctuation, koscielny2006long,kantelhardt2006long}. Although the considered rivers are located in the same region (namely, NEGOM), the autocorrelation analysis does not reveal features common to the different series analyzed. In particular, we observe slowly decay correlations, with different decorrelation times $\tau^{\star}$ ranging from 70 days to 250 days. In Fig.\,\ref{fig-power}(b) we show the logarithm of the power spectrum vs. the logarithm of the frequency for the same rivers: linear regions, corresponding to scaling regimes $P_{\kappa}(\nu)\sim \nu^{\beta_{k}}$ are evident, and can be characterized by the scaling exponents $\beta_{\kappa}$. Scaling regions of different length are present on different frequency ranges. In order to perform a simple comparison among the considered rivers, we consider only three particular frequency ranges and, for each range separately, we estimate the scaling exponents $\beta_{\kappa}$ for each river by considering linear regions of the same length. In Fig.\,\ref{fig-power}(c), we show the rank-ordered scaling exponents $\beta_{\kappa}$, extracted from the slopes of the linear regions in Fig.\,\ref{fig-power}(b), for $\log(\nu/\text{day}^{-1})\in[-2,-1]$, $[-3,-2]$ and $[-4,-3]$. The main result in the frequency domain is that, in general, time series measurements for each river flow are correlated, although they are characterized by several scaling regimes with exponents ranging from about -0.5 to about -5.0, and corresponding, in principle, to dynamical systems of very different nature. 

A more advanced approach to the study of correlations in time series involves the detrended fluctuation analysis (DFA) \cite{peng1994mosaic} or its generalized version, the multifractal DFA (MDFA) \cite{kantelhardt2002multifractal}. The MDFA is a powerful tool, strictly related to standard multifractal analysis,  allowing i) to explore the scaling relationship between detrended fluctuations and time as a function of a real variable $\gamma$, and ii) to characterize correlations by scaling exponents that generalize the Hurst parameter, the quantity widely used to quantify the amount of persistence or anti-persistence in time series. The Hurst parameter $\mathcal{H}_{\kappa}$, obtained from MDFA when $\gamma=2$, is strictly related to scaling exponents $\beta_{\kappa}$, obtained from spectral analysis, by $-\beta_{\kappa}=2\mathcal{H}_{\kappa}-1$ \cite{heneghan2000establishing}.

An accurate estimation of the Hurst parameter, through both MDFA and wavelet techniques, for several rivers around the world has been recently reported \cite{koscielny2006long,kantelhardt2006long,livina2007temporal}. The analysis put in evidence that detrended fluctuations are characterized by different values of the Hurst parameter. In particular, $\mathcal{H}$ varies from river to river, ranging from 0.55 to 0.95, underlying the long-term persistence of river flow dynamics on time scales from weeks to decades but no universal behaviour \cite{koscielny2006long}. Moreover, the multifractal analysis revealed a self-affine scaling behaviour and a universal function, derived from a generalization of the multiplicative random cascade model, has been found to describe the multifractal spectra of all rivers with extraordinary precision \cite{koscielny2006long,livina2007temporal}. River flow fluctuations should be strongly dependent on rainfall dynamics, and therefore multifractal cascade models should be the best candidates for their characterization \cite{tessier1996multifractal}. However, it has been recently shown that precipitations might be not responsible for both the long-term correlations observed in river flow data and the appropriateness of multiplicative cascades for their modelling \cite{kantelhardt2006long}. Although such results represent an important advance in understanding the dynamics of river flow fluctuations, a universal scaling behaviour in the time domain is still missing.

The goal of our spectral study is to show the great variance in the scaling exponents $\beta_{\kappa}$ estimated from three different frequency intervals and corresponding to time scales ranging from few days to few weeks. In order to compare our results with those reported in literature, for instance in Tab.\,1 of Ref.\,\cite{koscielny2006long}, we estimate the power spectrum of deseasonalized time series and the corresponding values of $\beta_{\kappa}$ for $-7.6<\log(\nu/\text{day}^{-1})<-4.5$ in the frequency domain or, equivalently, for $90<n/\text{day}<2000$ in the time domain. We have chosen this interval because it is compatible with the interval where Hurst parameters, reported in Ref.\,\cite{koscielny2006long}, have been determined for several rivers around the world. For instance, the time series for Thames river at Kingston ($\kappa=20$) is common to both analyses: in this case we obtain $\beta_{20}=-0.70$, corresponding to the value $\mathcal{H}_{20}=0.85\pm0.08$ for the Hurst parameter, in good agreement with the value $h(2)=0.8$ reported in literature \cite{koscielny2006long}. However, in this Letter, we are not interested in a direct comparison between our results and those ones obtained from MDFA \cite{koscielny2006long,livina2007temporal}, but we are interested in investigating the existence of a dynamics common to all rivers in the time domain, with no regards for climate, hydrological characteristics or geographical location.

To better capture the universal features of the underlying dynamics we adopt a more general approach, originally developed for the investigation of turbulent flows at different time scales \cite{castaing1990velocity}. 
It consists in analyzing a set of $N$ new time series, each constructed as the difference between the original time series, locally detrended to remove nonstationarity, and a copy of it delayed by $n$ ($n=1,2,...,N$) days. The detrending procedure is as follows. Each time series is divided into sliding segments of size $2n$ and, for each segment, the trend is obtained from the best $\mu-$th order polynomial $P_{\kappa}(t)=\alpha_{0}+\alpha_{1}t+\alpha_{2}t^{2}+...+\alpha_{\mu}t^{\mu}$ fitting the data, where the value of $\mu$ is varied from 0 up to 20. We have verified that for the time scales of interests to our study, smaller than six years, the 20th order is enough to safely remove seasonal trends. As for the order $\mu$ of the polynomial model that best approximates the original time series $q_{\kappa}(t)$, we choose the one that minimizes the Akaike information criterion \cite{Akaike74}, preventing over-fitting. The resulting time series of the difference between the data $q_{\kappa}(t)$ and the trend $P_{\kappa}(t)$ is then used for the analysis. We indicate by $Q_{\kappa}(t)=q_{\kappa}(t)-P_{\kappa}(t)$ the detrended time series obtained from such a procedure. It is worth remarking that data has not been deseasonalized before applying this procedure, because such a general approach is able to remove any trend, including the seasonal one, by construction. However, we have verified that pre-processing the data by the standard deseasonalizing technique has no impact on the final result. Let $\delta_{n} Q_{\kappa}(t)=Q_{\kappa}(t+n)-Q_{\kappa}(t)$ be the time series of increments, at the time scale $n$, for measurements at the $\kappa-$th station, and let $\langle Q_{n\kappa}\rangle$ and $\sigma_{n\kappa}^{2}$ be the corresponding sample mean and variance, respectively. In the following we indicate by $\Delta_{n} Q_{\kappa}(t)=\delta_{n} Q_{\kappa}(t)-\langle Q_{n\kappa}\rangle$ the time series of increments shifted to zero mean. As an example, in Fig.\,\ref{fig-river}(b)  we report the time series of the standardized increment $\Delta_{n} Q_{\kappa}(t)/\sigma_{n\kappa}$ obtained for the Alabama river when we adopt a time delay of $n=2$ days. 

\begin{figure}[!t]
  \begin{center}
      \includegraphics[angle=0, scale=0.3]{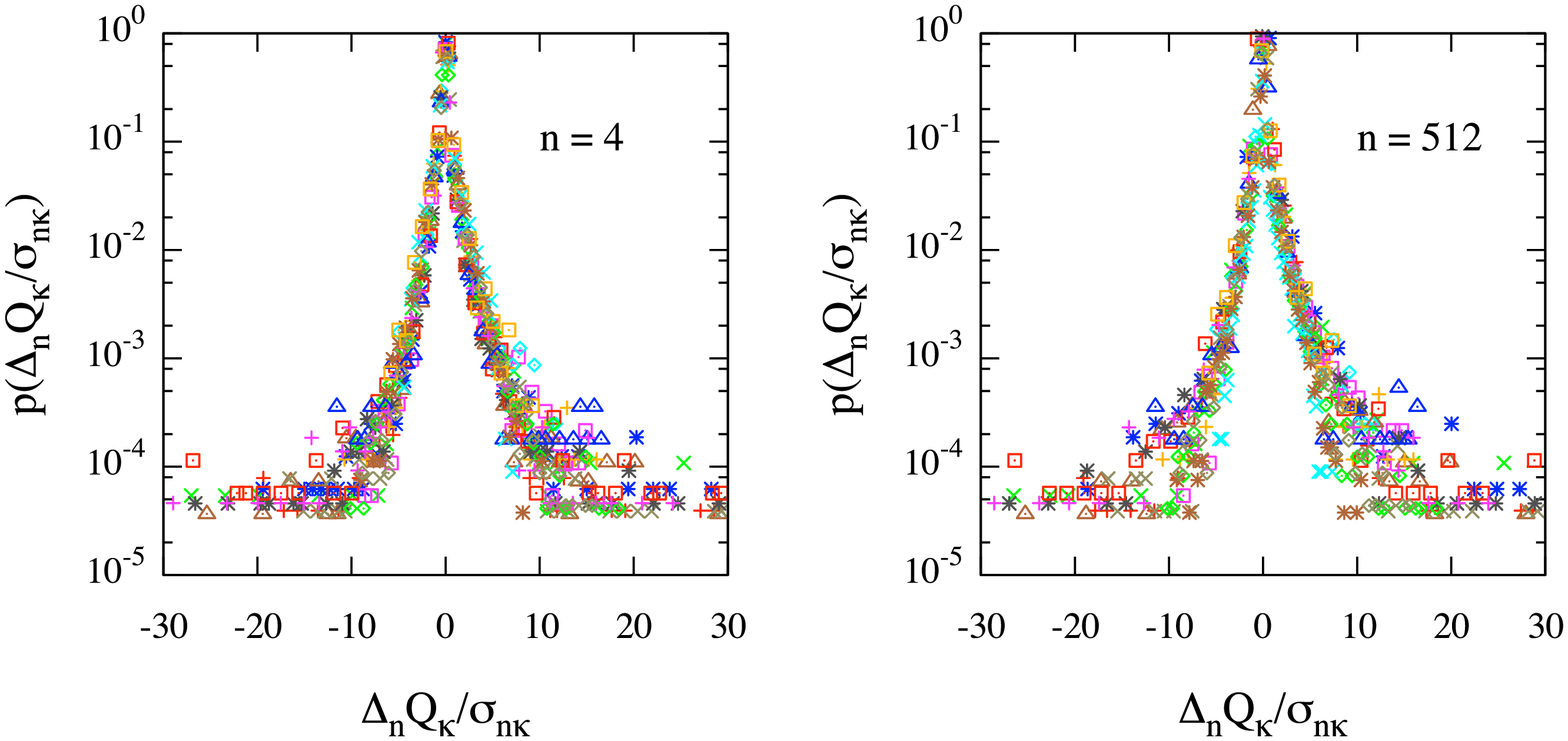}
   \caption{{Probability density of increments, rescaled to standard deviations $\sigma_{n\kappa}$, for the daily discharge as measured at the $\kappa-$th station ($\kappa=1,2,...,17$) in Tab.\,\ref{tab-rivers}. Two representative distributions, for $n=4$ days (left panel) and $n=512$ days (right panel), are shown.}}
    \label{fig-all-turb}
  \end{center}
  
  \begin{center}
     	\includegraphics[angle=0, scale=0.3]{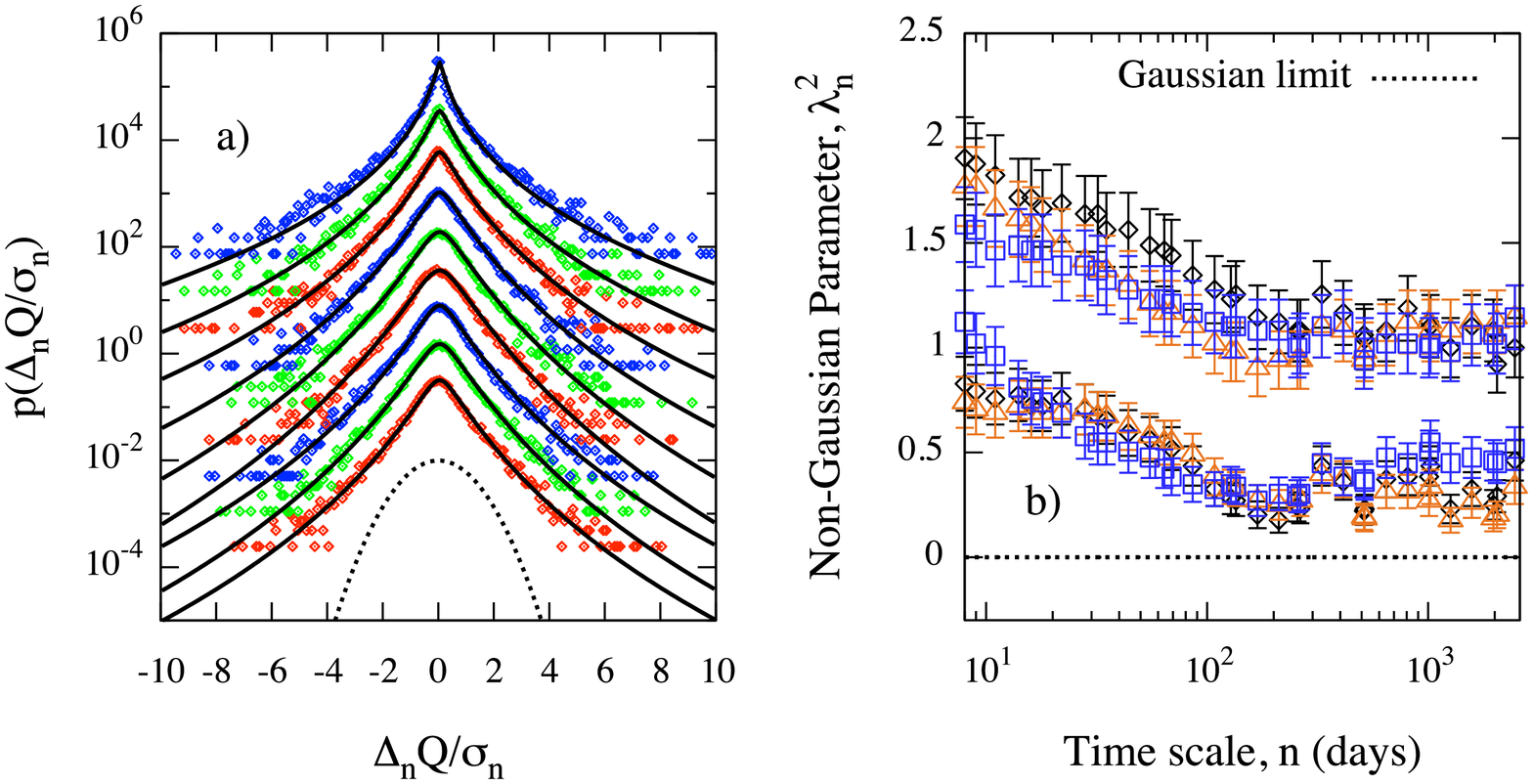}
      \caption{{Deformation of the PDF of standardized increments across scales. (\emph{a}) Vertically shifted PDFs of the Fish Creek river are shown for different time scales: $n=8, 16, 32, 64, 128, 256, 512, 1024$ and $2048$ days (from top to bottom), together with the corresponding Castaing fitting curves (solid lines). A Gaussian distribution is also shown for reference (dotted line) (\emph{b}) Scale dependence of the non-Gaussian parameter $\lambda_{n}^{2}$ vs. $\log n$ for Barwon ($\kappa=34$, upper diamonds), Inabòn ($\kappa=19$, upper squares), Tan-Shui ($\kappa=28$, upper triangles), Tombigbee ($\kappa=3$, lower diamonds), Verdalsvassdraget ($\kappa=24$, lower squares) and Thames ($\kappa=20$, lower triangles).}}    
      \label{fig-twocat}
  \end{center}

\end{figure}

In Fig.\,\ref{fig-all-turb} we show the PDFs of standardized increments for the stations in the NEGOM region reported in Tab.\,\ref{tab-rivers} ($\kappa=1,2,...,17$), and for two different values of time delay, namely $n=4$ days (left panel) and $n=512$ days (right panel). Strikingly the curves for different rivers all collapse into the same non-Gaussian distribution. Moreover, we have verified that for any value of the polynomial order $\mu$, ranging from 13 up to 250, results do not change. Such a result was expected. In fact, the largest time scale we consider is $n=2048$ days, corresponding to about six seasons: the procedure requires at most $\mu=12$ to deseasonalize segments of length $2n$, namely twelve seasons. In addition to this, we find that the same universal behavior holds at any time scale, from days to years. Similar anomalous behaviors have been observed in several other real-world processes of different nature, such as in the healthy human heart rate \cite{kiyono2004critical, kiyono2005phase}, in stock-price fluctuations \cite{mantegna1995scaling, kiyono2006criticality}, in human behavioral organization \cite{nakamura2007universal} and in seismic events \cite{manshour2009turbulencelike}.

\begin{figure}
  \begin{center}
      \includegraphics[angle=0, scale=0.3]{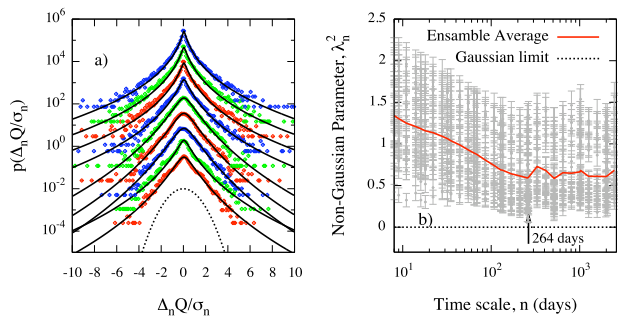}
   \caption{{Deformation of the PDF of standardized increments across scales. (\emph{a}) Same as in Fig.\,\ref{fig-twocat}(a) in the case of the Tombigbee river. (\emph{b}) Scale dependence of the non-Gaussian parameter $\lambda_{n}^{2}$ vs. $\log n$ for all time series in Tab.\,\ref{tab-rivers}, as well as the ensemble average for each $n$.}}
    \label{fig-castaing}
  \end{center}
\end{figure}

Models involving (eventually periodic) autoregressive approaches, provide Gaussian PDFs of delayed increments and thus fail to correctly reproduce the data at any time scale. Conversely, heavy-tailed non-Gaussian PDFs emerge, for instance, from random multiplicative processes \cite{castaing1990velocity, takayasu1997stable}. In this case, the simplest approach is to consider two indipendent Gaussian distributions, $F_{n}$ and $G_{n}$, depending on the time scale $n$, with zero mean and variances respectively equal to $\sigma_{n}^{2}$ and $\lambda_{n}^{2}$. If we define the increments $\Delta_{n}Z(t)$ as
\begin{eqnarray}
\label{def-Z}
\Delta_{n} Z(t) = \xi_{n}(t)e^{\omega_{n}(t)},
\end{eqnarray}
where  $\xi_{n}(t)\sim F_{n}(0,\sigma_{n}^{2})$ and $\omega_{n}(t)\sim G_{n}(0,\lambda_{n}^{2})$, it is possible to prove that the PDF is the Castaing function
\begin{eqnarray}
\label{castaing-pdf}
p_{n}(\Delta_{n} Z)=\int_{0}^{\infty} F_{n}\(\frac{\Delta_{n} Z}{\sigma}\)\frac{1}{\sigma}G_{n}(\ln \sigma) \emph{d}(\ln \sigma),
\end{eqnarray}
originally introduced for a log-normal cascade model of fully developed turbulence \cite{castaing1990velocity}. If delayed fluctuations $\Delta_{n} Z$ are standardized, it can be shown that $\sigma_{n}^{2}=\exp(-\lambda_{n}^{2})$ and the function in Eq. (\ref{castaing-pdf}) depends just on one parameter, namely $\lambda_{n}$. The standard Gaussian distribution of delayed increments is attained when $\lambda_{n}$ approaches zero: for this reason $\lambda_{n}^{2}$ has been commonly adopted as a key parameter in quantifying the non-Gaussian behavior of fluctuations \cite{kiyono2004critical,kiyono2005phase,kiyono2006criticality,nakamura2007universal,manshour2009turbulencelike}. We have therefore fitted the PDF of delayed increments obtained from our detrended river flow data
with the function in Eq. (\ref{castaing-pdf}), to estimate the non-Gaussian parameter $\lambda^{2}_{n}$ at different time scales $n$. As an example, the result of the fit is reported in Fig.\,\ref{fig-twocat}(a) and \ref{fig-castaing}(a) for the cases of the Fish Creek and Tombigbee rivers, respectively. The non-Gaussian distribution of fluctuations is clear at every time scale and it is well fitted by a Castaing function with a value of $\lambda^{2}_{n}$ which decreases with $n$. We have repeated the fitting procedure for each of the time series in Tab.\,\ref{tab-rivers}. 

Apparently, rivers exhibit a common scaling with no regards for drainage area, length, elevation, average annual precipitation or geographical position. Previous studies have already shown, for instance, the independence on the basin area or the geographical location of parameters characterizing the scaling behaviour in river flow data \cite{gupta1994multiscaling,koscielny2006long}. As a representative example, in Fig.\,\ref{fig-twocat}(b) is shown the dependence of the non-Gaussian parameter $\lambda_{n}^{2}$ on the scale $n$ for some rivers characterized by very different climate, hydrological characteristics and location. For Barwon (Victoria, Australia), the average annual rainfall ranges from 2000 mm, in the south of the Barwon Basin, to 700 mm in the northern region of the basin, whereas for Inabòn (Ponce, Puerto Rico) the average annual rainfall reaches 2500 mm. Although these two rivers have very different length (160 km and 28 km, respectively) and drainage area (1269 km$^{2}$ and 25 km$^{2}$, respectively), they exhibit the same scaling behaviour: $\lambda_{n}^{2}$ linearly depends on time scale $n$ up to a certain horizon, where a crossover with a new regime appears. The same result is found for rivers corresponding to points in the bottom of Fig.\,\ref{fig-twocat}(b), closer to the Gaussian limit. In this case the average annual precipitation is rather different: for instance, 700 mm for the Thames (UK, Europe) and 1200-1500 mm for the Tombigbee (Alabama, USA).

In Fig.\,\ref{fig-castaing}(b) we report, all together, the non-Gaussian parameters found for the 34 different rivers, and their average, as a function of the logarithm of the time scale $n$. The scale dependence of $\lambda^{2}_{n}$ is a common feature of all time series analyzed.
Our results clearly indicate that $\lambda^{2}_{n}$ scales, for values of $n$ up to 264 days, as the logarithm of $n$, rather than as a power law, while for $n>264$ $\lambda^{2}_{n}$ takes a constant value. In the Kolmogorov-Obukhov theory of turbulence \cite{obukhov1962,kolmogorov1962} the non-Gaussian parameter
$\lambda^{2}_{n}$ is expected to be linear in $\log n$ \cite{chabaud1994transition}, whereas a power law is expected in the case of Castaing model. Therefore, the fluctuations observed in river flows have a strong analogy with hydrodynamic turbulence-like behavior, for temporal scales up to about 264 days. However, log-normal models are unable to explain the nearly constant values of $\lambda^{2}_{n}$, observed for $n>264$, which indicate scale invariance, and suggest a transition to a critical regime where the PDFs of fluctuations all perfectly collapse on the same non-Gaussian curve.
For $n>264$ days, correlations become persistent and strong fluctuations develop at any time scale, from the shortest scales above the horizon up to the largest ones. 

Our results are intriguing for different reasons. First, we have found a novel scaling in the time domain, common to all rivers, with no regards for their hydrological features or their location: such behaviour is peculiar of multiplicative cascade systems and breaks up above a certain horizon, where scale-invariance appears. Second, although the values of the non-Gaussian parameter are not the same for all rivers, for a fixed time scale $n$, it is possible to capture a single dynamics, as shown in Fig.\,\ref{fig-castaing}(b). The main difference among rivers is not in the shape of $\lambda^{2}_{n}$, constrained by the underlying dynamics, but in the divergence from the limit of Gaussian fluctuations. Roughly speaking, the river flow dynamics appears to be always the same, and rivers just differentiate because of the variance of non-Gaussian fluctuations they exhibit. It is worth remarking that multiplicative cascade models, suggested in recent studies \cite{koscielny2006long,kantelhardt2006long,livina2007temporal}, represent ideal candidates to explain our results. In fact, such models have been successfully adopted for modelling turbulent dynamics (see Ref.\,\cite{gao2007multiscale}, and Refs. therein, for a comprehensive review of multiplicative models for turbulence) and it has been recently shown that a single universal function, deduced from a generalized multiplicative model and depending just on two parameters, is able to reproduce the multifractal spectra of many rivers around the world \cite{koscielny2006long}. The generalized multiplicative model accounts for both the scaling in the structure function for time scales greater than several weeks \cite{koscielny2006long} and the deformation of the PDF of standardized increment across scales up to about 264 days. The scale-invariance observed above the horizon cannot be accounted for when only multiplicative cascade models are considered \cite{kiyono2006criticality}. Recent studies reported a crossover time of several weeks, above which daily runoffs are long-term correlated \cite{koscielny2006long,kantelhardt2006long}. Such crossover time is about half of the value we have found within the present study for the transition between two different regimes. However, the two results can not be directly compared because they have been obtained from the investigation of quite different features of river flow, the scaling in the structure function, in one case, and the scaling of non-Gaussian curves in our case.


Summing up, in this Letter we have found a novel scaling for river dynamics which relates the non-Gaussian fluctuations occurring in the flows at different time scales. Both the scaling behavior and the break up of the scaling are features common to all the rivers we have considered. Moreover, recent studies involving multifractal detrended fluctuation analysis and multifractal formalism show that multiplicative cascades, and their extended versions, are able to characterize a large number of rivers \cite{koscielny2006long,kantelhardt2006long,livina2007temporal}. The universality of our results supports the appropriateness of random multiplicative models, setting up a definitive framework for modelling flow dynamics, but suggest, at the same time, the need for a generalization that is able to reproduce both well-known results in literature and the novel scale-invariance above the critical horizon. 

The authors thank K. Kiyono for useful discussions on the multiscale fluctuation analysis and the anonymous referees for interesting comments and suggestions.

\addcontentsline{toc}{section}{References} 

\bibliographystyle{eplbib.bst} 
\bibliography{draft}

\end{document}